\documentclass[aip,graphicx]{revtex4-1}
\usepackage[draft]{todonotes}
\usepackage{subcaption}

\draft % marks overfull lines with a black rule on the right
\usepackage{graphicx}
\graphicspath{{figures/}}

\begin{document}

% Use the \preprint command to place your local institutional report number
% on the title page in preprint mode.
% Multiple \preprint commands are allowed.
\preprint{Preprint submitted to Journal of Renewable and Sustainable Energy
19 March 2016}

\title{Modeling the near-wake of a vertical-axis cross-flow turbine with 2-D and
    3-D RANS}

% repeat the \author .. \affiliation  etc. as needed
% \email, \thanks, \homepage, \altaffiliation all apply to the current author.
% Explanatory text should go in the []'s,
% actual e-mail address or url should go in the {}'s for \email and \homepage.
% Please use the appropriate macro for the type of information

% \affiliation command applies to all authors since the last \affiliation command.
% The \affiliation command should follow the other information.

\author{Peter Bachant}
\email[]{pete.bachant@gmail.com}
%\homepage[]{Your web page}
%\thanks{}
%\altaffiliation{}
\affiliation{Center for Ocean Renewable Energy, University of New Hampshire,
Durham, NH}

\author{Martin Wosnik}
%\email[]{}
%\homepage[]{Your web page}
%\thanks{}
%\altaffiliation{}
\affiliation{Center for Ocean Renewable Energy, University of New Hampshire,
Durham, NH}

% Collaboration name, if desired (requires use of superscriptaddress option in
% \documentclass).
% \noaffiliation is required (may also be used with the \author command).
%\collaboration{}
%\noaffiliation

\date{\today}

\begin{abstract}
    The near-wake of a vertical-axis cross-flow turbine (CFT) was modeled
    numerically via blade-resolved $k$--$\omega$ SST and Spalart--Allmaras RANS
    models in two and three dimensions. Results for each case are compared with
    experimental measurements of the turbine shaft power, overall drag, mean
    velocity, turbulence kinetic energy, and momentum transport terms in the
    near-wake at one diameter downstream. It was shown that 2-D simulations
    overpredict turbine loading and do not resolve mean vertical momentum
    transport, which plays an important role in the near-wake's momentum
    balance. The 3-D simulations fared better at predicting performance, with
    the Spalart--Allmaras model predictions being closest to the experiments.
    The SST model more accurately predicted the turbulence kinetic energy while
    the Spalart--Allmaras model more closely matched the momentum transport
    terms in the near-wake. These results show the potential of blade-resolved
    RANS as a design tool and a way to ``extrapolate'' experimental flow field
    measurements.
\end{abstract}

% insert suggested PACS numbers in braces on next line

\maketitle %\maketitle must follow title, authors, abstract and \pacs
% \listoftodos

\section{Introduction}

In the pursuit of a fully sustainable energy profile, cross-flow (often
vertical-axis) turbines (CFTs) can still play a role. Despite extensive research
and development in the 1970s through the early 1990s by groups like Sandia
National Labs in the US \cite{Sutherland2012} and the National Research Council
in Canada \cite{Para2002}, the vertical-axis cross-flow Darrieus turbines were
all but abandoned for large scale commercial terrestrial wind power. Today,
however, with the development of marine hydrokinetic (MHK) energy devices, CFTs
are at the forefront, with the first grid-connected commercial device being a
horizontal-axis CFT manufactured and installed by ORPC in Cobscook Bay, Maine
\cite{ORPC2012}. CFTs are also being examined for small-scale onshore wind in
urban areas \cite{Lott2015}, offshore floating wind farms \cite{Paulsen2011,
Sandia2012}, and onshore wind farms where power output per unit land area is of
interest \cite{Dabiri2011}.

Despite the development of many simple engineering models based on blade element
momentum or vortex methods, it remains difficult to predict the performance of
cross-flow turbines for all cases---namely when solidity, or blade
chord-to-radius ratio is high---a common characteristic of smaller rotors, and
those designed for marine hydrokinetic (MHK) applications. With computing power
becoming evermore available and affordable, CFD based on the Reynolds-averaged
Navier--Stokes (RANS) equations has become an attractive method for predicting
the performance of cross-flow turbines. Since a direct numerical simulation
(DNS) of all scales at realistic Reynolds numbers is not currently feasible,
turbulence be modeled. However, with an appropriate turbulence model,
blade-resolved RANS presents a more physically realistic first principles based
approach versus simpler models based on momentum theory or potential flow.
However, CFD can be computationally expensive when done in three dimensions,
which may be necessary in some cases.

There are many examples in the literature of 2-D cross-flow turbine simulations
with widely varying results. Balduzzi et al.~\cite{Balduzzi2016} provides
a summary of recent efforts and an attempt to standardize a methodology for
using Reynolds-averaged Navier--Stokes (RANS) to correctly predict performance
of a 2-D CFT. Howell et al.~\cite{Howell2010}, performed both 2-D and 3-D
simulations of a high solidity cross-flow turbine using a $k$--$\epsilon$
renormalization group (RNG) turbulence model. The results from the 2-D
simulations over-predicted power coefficient, while the 3-D case matched well
with wind tunnel measurements near the tip speed ratio of maximum power. In
general, 3-D simulations are less common, but have begun to appear more
frequently recently---a testament to the progress towards higher computing
power.

An overview of 3-D blade-resolved cross-flow turbine simulations reported in the
literature is presented in Table~\ref{tab:cfd-refs}. The $k$--$\omega$ SST RANS
turbulence model is shown to be a popular choice due to its success in
predicting flows with adverse pressure gradients and
separation~\cite{Menter2003}. Higher fidelity methods that resolve the large
scales of turbulence, such as large eddy and detached eddy simulation have also
been used. Note that for all the studies listed, model validation was performed
for either performance or wake predictions---not both---and in some case omitted
entirely.

\begin{table}
    \centering
    \begin{tabular}{c|c|c|c}
        Author & Turbulence modeling & Perf. val. & Wake val. \\
        \hline
        Alaimo et al.~\cite{Alaimo2015} & $k$--$\epsilon$ RANS & N/A & N/A \\
        Marsh et al.~\cite{Marsh2015} & SST RANS & Reference~\cite{Rawlings2008} & N/A \\
        Orlandi et al.~\cite{Orlandi2015} & SST RANS & References~\cite{Akins1989,Mertens2003} & N/A \\
        Lam \& Peng~\cite{Lam2016} & SST RANS \& IDDES & N/A & Reference~\cite{Tescione2014} \\
        Nini et al.~\cite{Nini2014} & Spalart--Allmaras RANS & N/A & Reference~\cite{Battisti2011} \\
        Boudreau \& Dumas~\cite{Boudreau2015} & Spalart--Allmaras DDES & N/A & N/A \\
        Li et al.~\cite{Li2013} & SST RANS \& Smagorinsky--Lilly LES & Reference~\cite{McLaren2011} & N/A \\
        Howell et al.~\cite{Howell2010} & $k$--$\epsilon$ RNG RANS & Reference~\cite{Howell2010} & N/A
    \end{tabular}

    \caption{Selected 3-D blade-resolved cross-flow turbine simulations reported
        in the literature, turbulence modeling employed, and performance and/or wake
        studies used for validation. Note the Li et al. study used periodic boundary
        conditions and is technically considered 2.5-D.}

    \label{tab:cfd-refs}
\end{table}

Modeling the boundary layer flows on cross-flow turbine blades is essential to
predicting the blade loading. This flow condition presents a challenge due to
the dynamically changing inflow velocity and angle of attack---which often
exceeds static stall values and causes dynamic stall. Furthermore, the ability
to predict the occurrence and interdependence of boundary layer transition to
turbulence and separation can have dramatic influence on the blade loading and
therefore the predicted turbine power output. These challenges present
significant obstacles to the prospect of using CFD to replace wind tunnel or
tank testing of physical models.

To date, little computational work has been done to attempt to design arrays of
CFTs, despite their prospects for closer spacing compared with axial-flow
turbines (AFTs). For example, Araya et al.~\cite{Araya2014} modeled the flow
through a VAT array using ``leaky Rankine body'' potential flow singularities,
which was able to rank relative---though not absolute---performance of array
configurations. Goude and Agren~\cite{Goude2010} used a 2-D vortex method to
simulate a farm of cross-flow turbines, though this was not validated with
experiments. Durrani et al.~\cite{Durrani2011} used 2-D CFD to model a group of
cross-flow turbines, observing higher power output for a staggered
configuration, but also did not compare with experimental results. Giorgetti et
al.~\cite{Giorgetti2015} took a similar approach for 2-D array analysis using
turbine pairs inspired by Dabiri~\cite{Dabiri2011}, but again experimental
validation was not performed. Li and Calisal \cite{Li2010} used a 3-D vortex
line method to show mutually improved power output from two adjacent turbines,
though the simulations over-predicted the effects by approximately 5\% compared
with experiments. Antheaume et al.~\cite{Antheaume2008} used a blade element
approach coupled with a 3-D RANS solver to also show how close spacing can
improve power output of CFTs.

In this study we set out to model the performance and near-wake of the high
solidity University of New Hampshire Reference Vertical-Axis Turbine (UNH-RVAT)
using the Reynolds-averaged Navier--Stokes equations using the open-source
finite volume CFD package OpenFOAM, version 2.3.x. Though studies in the
literature generally focus on predicting the turbine loading and the local blade
boundary layer, we seek to model both this and the larger scale flow produced by
the rotor, i.e., the near-wake, which is of interest for this particular turbine
since it has been shown experimentally that the near-wake's momentum and energy
transport processes are dominated by vertical advection \cite{Bachant2015-JoT}.
It logically follows that a 2-D simulation, which omits the vertical dimension,
would not correctly predict wake recovery and turbine--turbine interaction.
However, it is of interest to determine how wrong a 2-D model may be, since the
lower computational cost of 2-D simulations is attractive.

We seek to validate 2-D and 3-D blade-resolved RANS models against the UNH-RVAT
mechanical power and near-wake measurements. If the blade loading and velocity
in the near-wake match well enough, the flow field can potentially be inspected
in greater detail, i.e., we may be able to make observations of quantities not
measured experimentally, e.g., pressure, and gain greater insight into where the
dominant flow structures originate. This will ultimately help develop and
evaluate low-order wake generator models for use in turbine array modeling. In
summary, the questions we hope to answer here are:

\begin{enumerate}
    \item Can 2-D RANS be used for individual turbine and/or array design?

    \item How accurately can 3-D RANS predict performance?

    \item Do the flow fields predicted by 3-D RANS match the experimental wake
    measurements well enough to provide insight for developing new low-order
    wake generators to represent CFTs?

    \item Does 3-D RANS realize the correct proportions of wake recovery
    mechanisms, i.e., are the 3-D blade-resolved results a good reference case
    or ``target'' for those a low-order model should produce?
\end{enumerate}

\section{Numerical setup}

\begin{figure}
    \centering

    \includegraphics[width=0.9\textwidth]{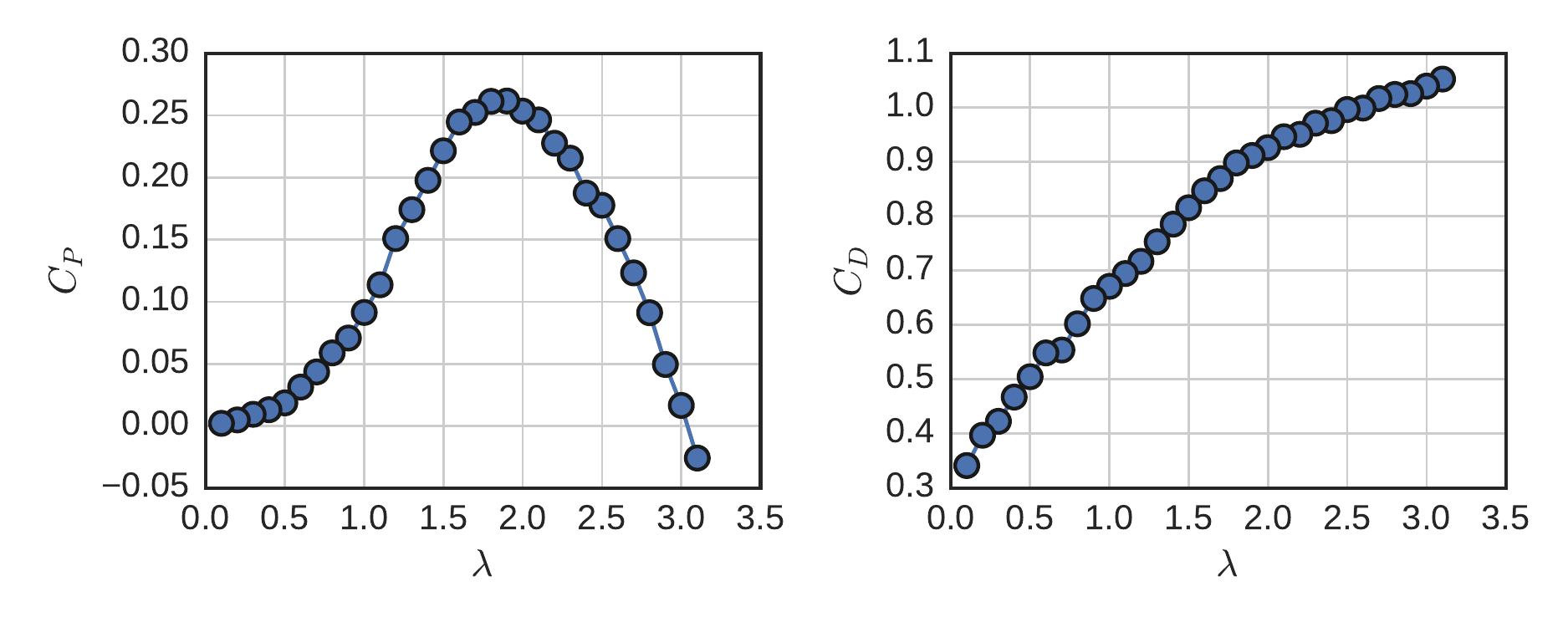}

    \caption{Mean rotor power (left) and drag (right) coefficient curves from
        the tow tank experiments\cite{Bachant2016-RVAT-Re-dep}.}

    \label{fig:exp-perf}
\end{figure}

In this study the UNH-RVAT baseline tow tank experiment was simulated using a
mean rotor tip speed ratio $\lambda=1.9$, which corresponds to the maximum
measured power coefficient; cf. Figure~\ref{fig:exp-perf}. The tow speed or free
stream velocity $U_\infty=1.0$ m/s gives a turbine diameter Reynolds number
$Re_D \approx 10^6$, which corresponds to the $Re$-independent state for both
performance and near-wake characteristics, as determined from previous
experimental measurements \cite{Bachant2014, Bachant2016-Energies}.

\begin{figure}
    \centering

    \includegraphics[clip, trim=0 1in 0 1in, width=0.8\textwidth]{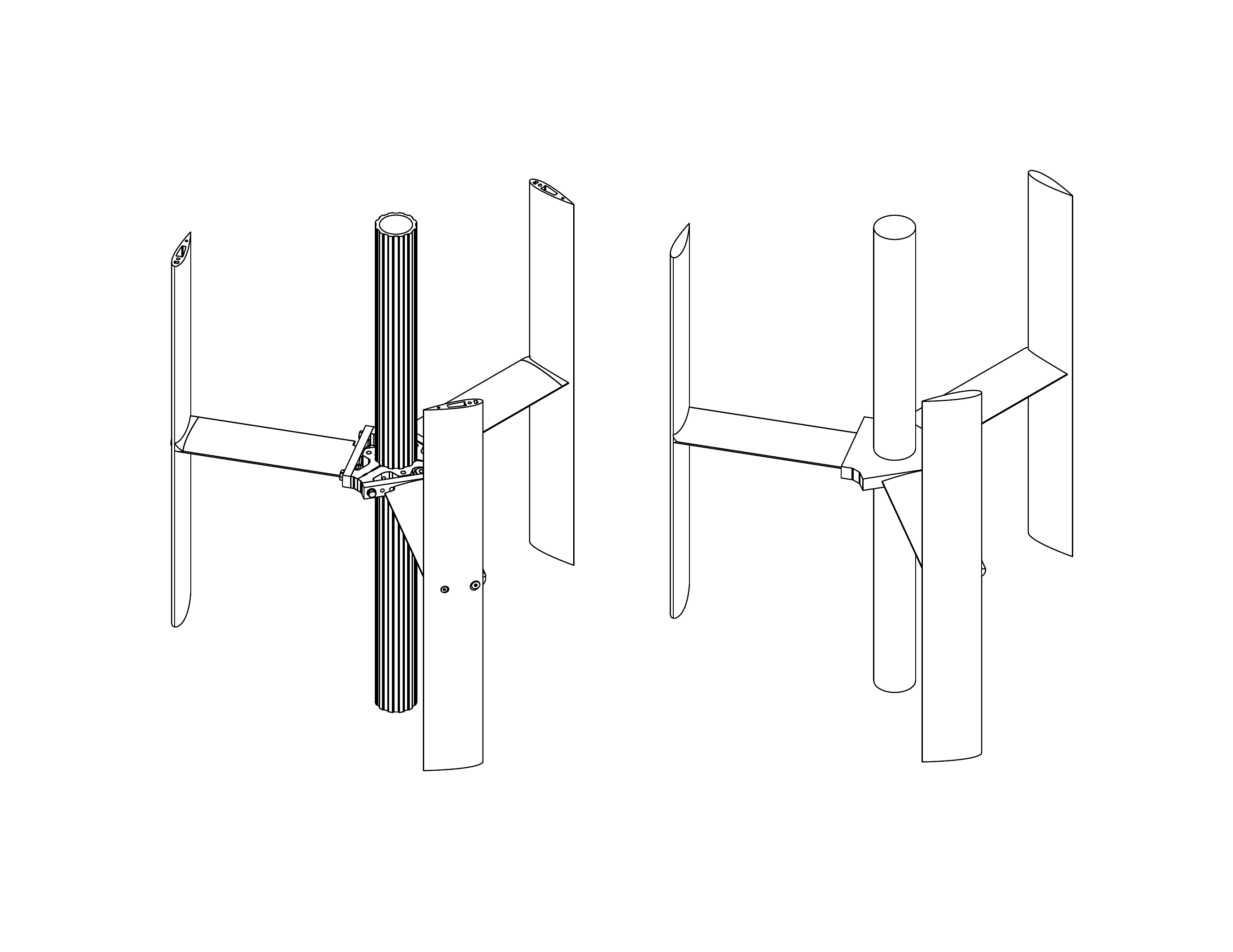}

    \caption{CAD drawings of the UNH-RVAT cross-flow turbine as designed (left)
        and cleaned for simulation (right).}

    \label{fig:cfd-cad}
\end{figure}

The turbines CAD geometry was prepared or ``cleaned'' for CFD by removing
details determined to be unnecessary, e.g., screw heads and axial shaft grooves,
which would complicate the meshing process and contribute very little to the
overall loading or flow modulation. A drawing of the physically and numerically
modeled geometries is presented in Figure~\ref{fig:cfd-cad}.

To close the RANS equations, two different turbulence models were used: Menter's
$k$--$\omega$ SST \cite{Menter1994} and the Spalart--Allmaras (SA) one equation
model \cite{Spalart1992}. Both closures use the eddy-viscosity approach---the SA
employing a single additional scalar transport equation for an
eddy-viscosity-like quantity $\tilde{\nu}$ and the SST solving equations for the
turbulence kinetic energy $k$ and specific dissipation $\omega$. The SST model
was chosen due to its prominence in the literature for simulating separating
flows, which we assumed to be present in the current problem in the form of
dynamic stall. The SA model was shown by Ferreira et al.~\cite{Ferreira2007} to
match experimental particle image velocimetry (PIV) results for a CFT in dynamic
stall, though this was a somewhat low Reynolds number case ($Re_c \approx 6
\times 10^4$). Further justification for using the SA model for this case comes
from Crivellini and D'Alessandro \cite{Crivellini2014}, where they successfully
modeled the laminar separation bubble and subsequent boundary layer transition
to turbulence at Reynolds numbers similar to those investigated here.

\subsection{Computational mesh}

\begin{figure}
    \centering

    \includegraphics[width=0.8\textwidth]{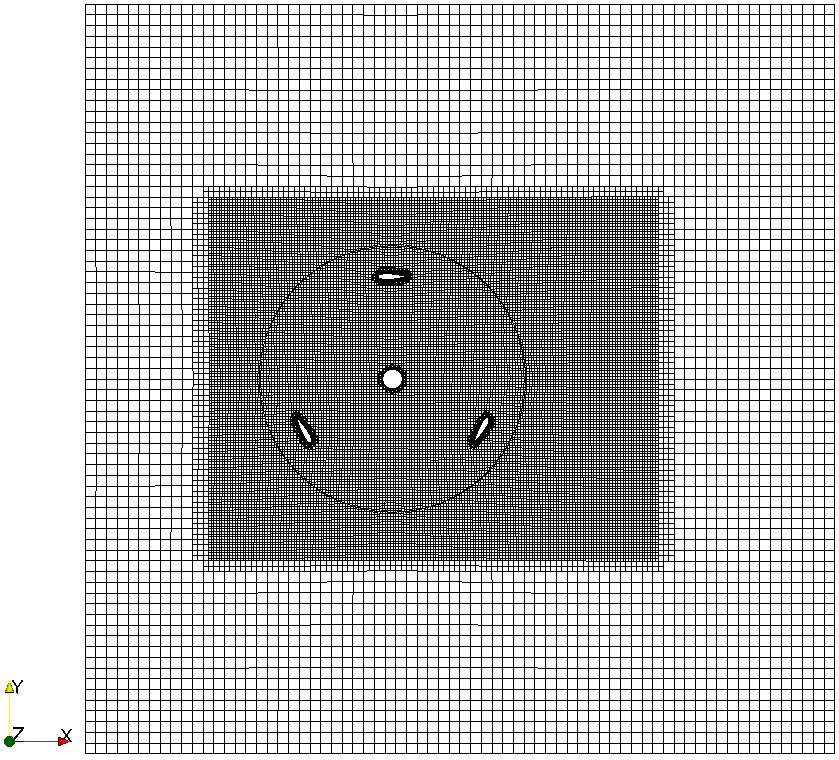}

    \caption{Overview of the 2-D computational mesh.}

    \label{fig:2d-br-mesh}
\end{figure}

\begin{figure}
    \centering

    \includegraphics[width=0.7\textwidth]{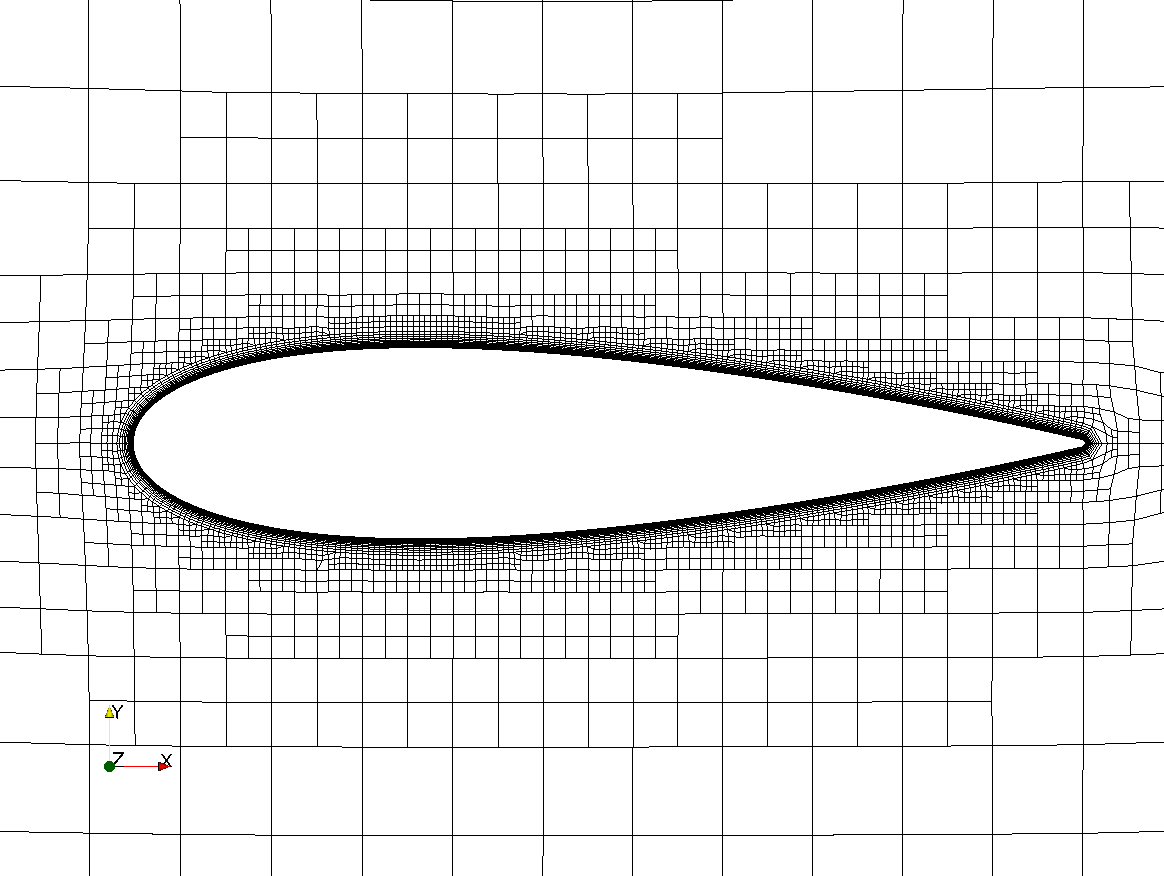}

    \caption{Detailed view of the 2-D computational mesh near the blades.}

    \label{fig:blade-mesh}
\end{figure}

The computational domain was a rectangular volume 3.66 m long, 3.66 m wide, and
2.44 m tall (for 3-D), with the turbine located 1.52 m from the inlet, and
centered vertically with a vertical axis, designed to match the tow tank
dimensions for comparison with previous experiments. The rotor geometry was
located at the center of a cylindrical sliding mesh interface, which rotated at
a mean tip speed ratio $\lambda=1.9$ with a sinusoidal oscillation at the blade
passage frequency---with and amplitude of 0.19 and the first peak at 1.4
radians---to mimic the slight deviation from the mean tip speed ratio observed
in the experiments. The 2-D mesh overview is shown in
Figure~\ref{fig:2d-br-mesh} and the blade mesh detail is shown in
Figure~\ref{fig:blade-mesh}.

Meshes were generated using OpenFOAM's blockMesh and snappyHexMesh utilities.
Mesh topology consists of a background hexahedral mesh, which is refined in all
three directions by a factor of 2 in a rectangular region containing the turbine
and near-wake (0.9 m upstream, 1.3 m downstream, $\pm 0.9$ m cross-stream, and
$\pm 0.8$ m vertically). Cells adjacent to the turbine shaft and struts are
refined by a factor of 4, while cells adjacent to the blades are refined by a
factor of 6. To capture the boundary layer, 20 layers were added next to the
blades with an expansion ratio of 1.2. Overall mesh refinement is controlled by
a single parameter---the number of cells in the streamwise direction, $N_x$.

\subsection{Solver}

Simulations were run using OpenFOAM's pimpleDyMFoam solver, which uses a hybrid
PISO-SIMPLE algorithm for pressure-velocity coupling and is compatible with
dynamic meshes. An Euler scheme was used to advance the simulation forward in
time. The case files required to replicate the simulations are available from
\cite{Bachant2016-UNH-RVAT-2D-OpenFOAM-SST, Bachant2016-UNH-RVAT-2D-OpenFOAM-SA,
Bachant2016-UNH-RVAT-3D-OpenFOAM-SST, Bachant2016-UNH-RVAT-3D-OpenFOAM-SA}.

\subsection{Initial and boundary conditions}

Initial and boundary conditions were set to match those of the tow tank as well
as possible. The velocity at the inlet, bottom, and side walls was fixed to 1
m/s to match the tow tank case, while the top boundary condition was a slip
velocity condition. Pressure was held fixed at the outlet while a zero-gradient
boundary condition was applied at the inlet (a typical
velocity-inlet/pressure-outlet case). Note that in 2-D the top and bottom
boundary conditions are ``empty,'' which is an OpenFOAM convention to indicate
two-dimensionality.

Since the mesh was only refined next to the blade surfaces in order to resolve
the boundary layer profile, no wall functions were used. However, wall functions
were used for the other turbine surfaces, i.e., struts, hub, and shaft, and for
the tank sidewalls, bottom, and top.

\section{Model verification}

\begin{figure}
    \centering

    \includegraphics[width=0.9\textwidth]{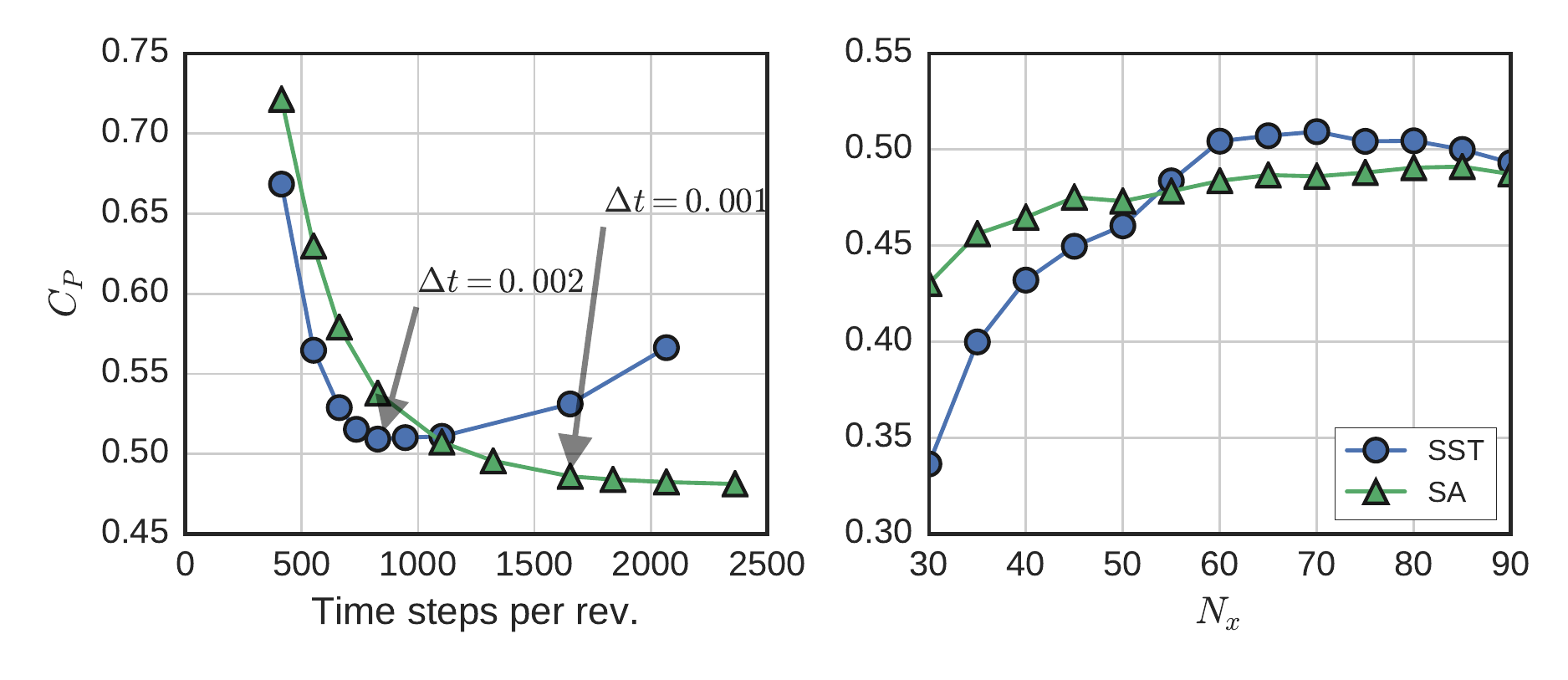}

    \caption{Time step (left) and grid size (right) dependence for the 2-D case
    with both the SST and SA turbulence models. Time step dependence was carried
    out with $N_x=70$ and grid size dependence with the time steps annotated for
    each turbulence model.}

    \label{fig:2d-br-verification}
\end{figure}

Both the $k$--$\omega$ SST and Spalart--Allmaras RANS model cases were verified
for convergence of the turbine mean power coefficient with respect to grid
spacing and time step using 2-D domains. The grid topology was fixed, but the
number of cells per domain length was scaled proportionally, maintaining the
same background mesh cell aspect ratio. Results for this parameter sweep are
shown in Figure~\ref{fig:2d-br-verification}, from which the final number of
streamwise grid points $N_x = 70$ was chosen, which resulted in a nondimensional
wall distance $y^+ = u^* y / \nu \sim 1$, where $u^*$ is the friction velocity,
defined as $u^*=\sqrt{\tau_w / \rho}$, where $\tau_w$ is the wall shear stress.
This mesh resolution gave a total cell count of approximately $5 \times 10^4$
for the 2-D cases and 16 million for the 3-D case. Sensitivity to domain length
was assessed as well, showing a 2\% increase in mean power coefficient for the
2-D SST case with the domain extended an additional two rotor diameters
downstream.

Time step dependence was evaluated using the 2-D $N_x=70$ grid, the results from
which are shown in Figure~\ref{fig:2d-br-verification}. It was seen that the
Spalart--Allmaras model converged well with decreasing time step, leading to a
final time step of 0.001 s. The results from the SST model show a local minimum
at $\Delta t = 0.002$ s, with some divergence for smaller time steps. The local
minimum was chosen as the final time step to run the simulations. Note that the
SST model's convergence behavior may be due to its specific implementation in
OpenFOAM, and not indicative of the nature of the model equations. Verification
studies for CFTs with this level of detail in the literature are not common,
though the final time step is comparable to others \cite{Balduzzi2016}.

\section{Results}

Turbine operation in all cases was simulated for 10 seconds, or approximately
six rotor revolutions. Computations for the 3-D cases were run on 192 processes
(24 nodes $\times$ 8 cores each on Sandia National Labs' Red Mesa high
performance computing cluster) and took on the order of 1,000 CPU hours per
second of simulated time. The 2-D simulations were run on a single processor and
took on the order of one CPU hour per second.

Performance statistics were computed after the first revolution onward and flow
statistics were calculated over the time interval spanning $5$--$10$ seconds, or
approximately three rotor revolutions, from time series downsampled to 50 Hz. It
is assumed that the downsampling frequency is high enough above the blade
passage frequency such that differences from the variance (for computing
turbulence statistics) in the original velocity will be negligible.

\subsection{Performance prediction}

\begin{figure}
    \centering

    \includegraphics[width=0.95\textwidth]{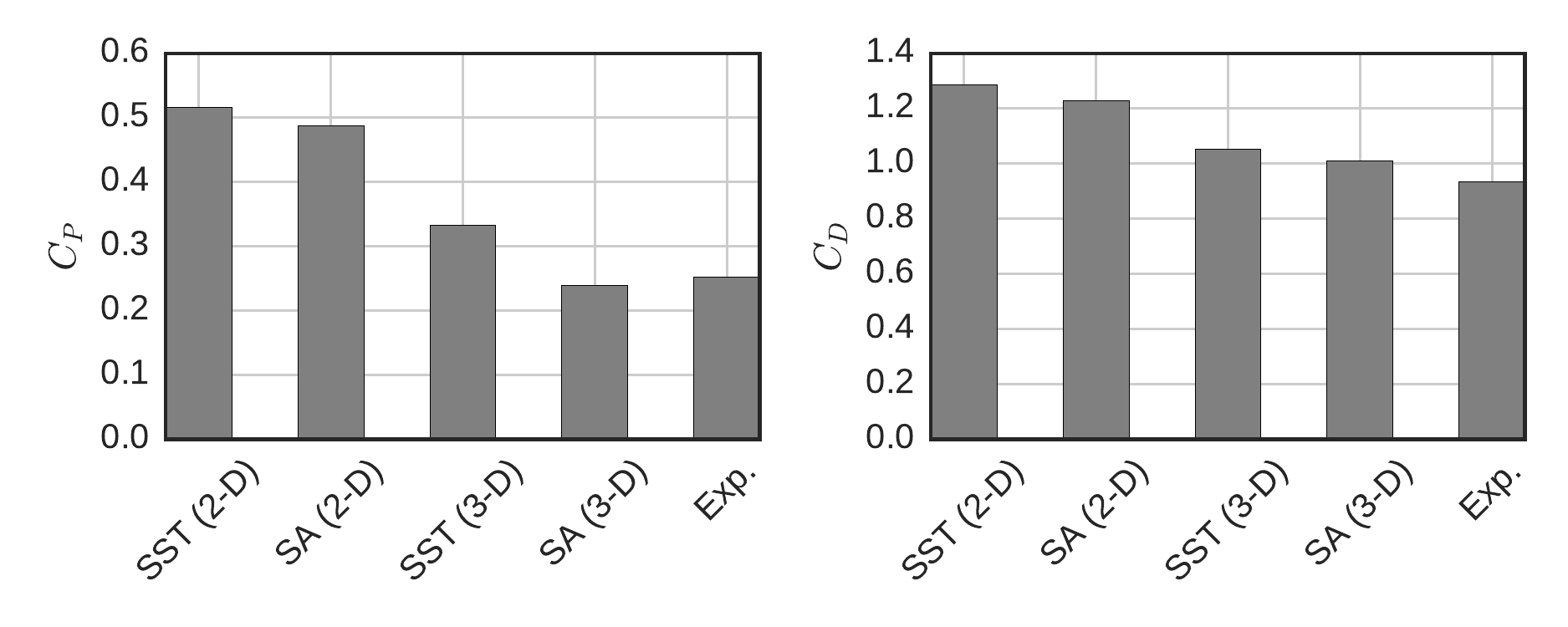}

    \caption{Power (left) and drag (right) coefficient predictions from
        experiments and each numerical model.}

    \label{fig:br-cfd-perf-bar-chart}
\end{figure}

Predictions for both the mean rotor power and drag coefficients are shown in
Figure~\ref{fig:br-cfd-perf-bar-chart}. In general, the 2-D CFD cases both
significantly overpredict turbine loading and therefore mechanical power output,
which is due to their increased blockage ratio, unresolved blade end effects,
and lack of blade support struts.

The 3-D simulations matched the experimental measurements more closely. The
Spalart--Allmaras model's mean power coefficient was 0.24 while the SST's was
0.33, which represent a 6\% under- and 30\% overprediction, respectively,
compared with the experiments. Some of the apparent overprediction of rotor drag
coefficient could also be an effect of the experimental procedure, where the
``tare drag'' from the turbine mounting structure was measured without a turbine
installed, then subtracted in post-processing. With the turbine installed, flow
past the mounting frame will be higher due to blockage, meaning the tare drag
would be underestimated, and the CFD results for $C_D$ should be slightly higher
than the experimental measurements.

\subsection{Wake characteristics}

\begin{figure}
    \centering

    \includegraphics[width=0.75\textwidth]{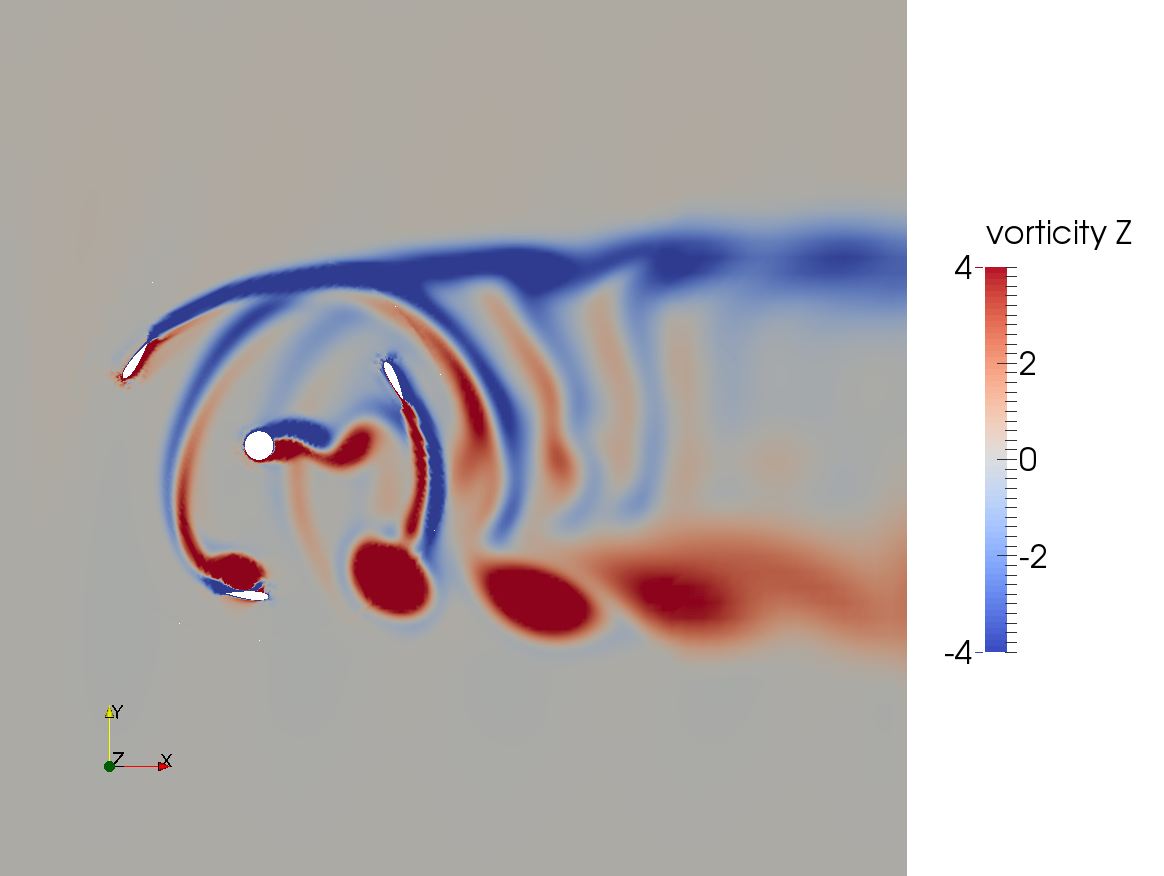}

    \caption{Instantaneous vorticity contours (at $t=9.64$ s) computed for the
        2-D Spalart--Allmaras case.}

    \label{fig:br-vorticity-2d}
\end{figure}

\begin{figure}
    \centering

    \includegraphics[width=0.75\textwidth]{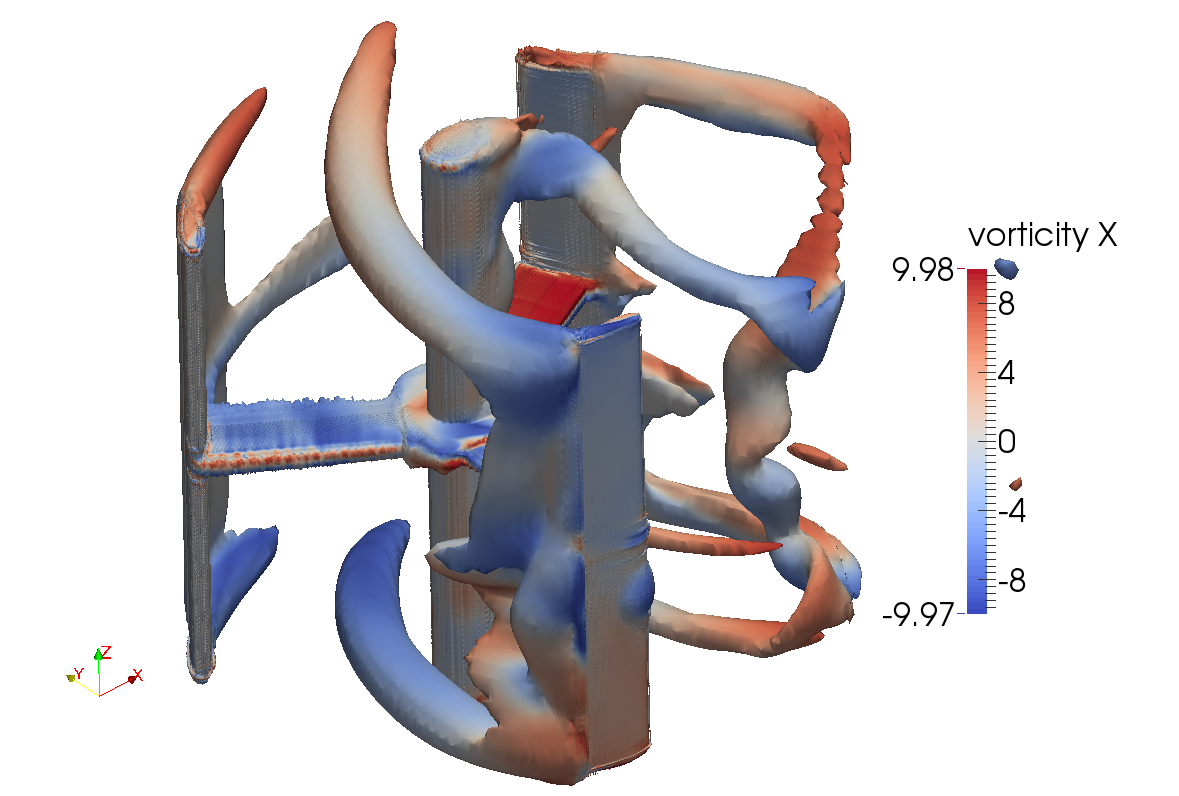}

    \caption{Iso-vorticity contours (at $t=9.64$ s) colored by the streamwise
        component of vorticity for the 3-D Spalart--Allmaras case.}

    \label{fig:br-vorticity-3d}
\end{figure}

Visualizations of the complex vorticity field generated by the turbine are
presented for the 2-D and 3-D Spalart--Allmaras cases in
Figure~\ref{fig:br-vorticity-2d} and Figure~\ref{fig:br-vorticity-3d},
respectively. It can be seen how the upstream blade---as it turns back into the
streamwise direction---is shedding a large amount of spanwise vorticity due to
the separated flow. In the 3-D case, strong tip vortices are also present, which
trace the ``contracting'' wake flow on the $-y$ side of the turbine associated
with the induced vertical velocity field. The 3-D dynamic stall vortex also
shows asymmetry about the $x$--$y$ mid-rotor plane; once again highlighting the
importance of three-dimensional effects on wake dynamics.

\begin{figure}
    \centering

    \includegraphics[width=0.95\textwidth]{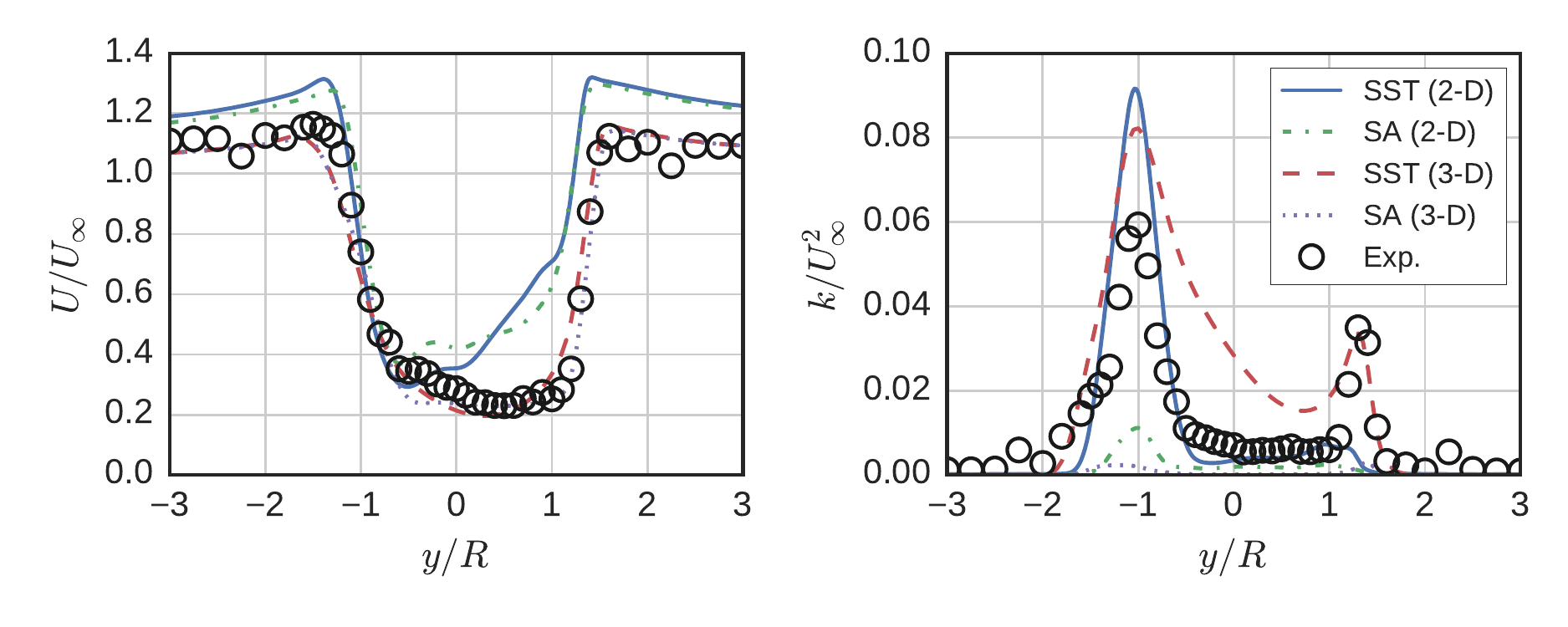}

    \caption{Mean velocity (left) and turbulence kinetic energy (right) profiles
        at $x/D=1$ from 2-D simulations, 3-D simulations ($z/H=0$), and experiments
        \cite{Bachant2015-JoT}.}

    \label{fig:br-cfd-profiles}
\end{figure}

Mean velocity profiles at one turbine diameter downstream are shown in
Figure~\ref{fig:br-cfd-profiles}. The 2-D results suffer from a blockage
mismatch, i.e., keeping the proximity of the walls constant increases the
blockage ratio. The 3-D results, however, show good agreement with the
experiments.

Turbulence kinetic energy profiles are also shown in
Figure~\ref{fig:br-cfd-profiles}. The turbulence kinetic energy for the unsteady
RANS models was calculated as
\begin{equation}
    k = k_{\mathrm{RA}} + \frac{1}{2} \left(
    \overline{U^\prime}^2 +
    \overline{V^\prime}^2 +
    \overline{W^\prime}^2 \right),
    \label{eq:k}
\end{equation}
where $U^\prime = U - \overline{U}$ (the resolved velocity fluctuations) and
$k_{\mathrm{RA}}$ is the kinetic energy calculated by the turbulence model,
which is zero for the SA model.

Both Spalart--Allmaras cases did a poor job predicting the turbulence kinetic
energy in the flow, since it must be resolved as variance in the velocity field.
The 2-D SST model did a good job predicting the peak in $k$ at $y/R=-1$, though
is missing the smaller peak at $y/R=-1$. This is once again likely due to
blockage issues, where local tip speed ratio is decreased, increasing the
blades' instantaneous angle of attack at this location on the downstream
passage. In contrast, the 3-D SST model predicts the $+y$ peak in turbulence
kinetic energy very well, though the $-y$ peak magnitude is overpredicted by
about 30\%. We also see some smearing of $k$ across the center of the rotor,
which is likely due to exaggerated levels of the turbulent eddy viscosity.

\subsubsection{Mean velocity in three dimensions}

\begin{figure}
    \centering
    \begin{subfigure}[b]{\textwidth}
        \centering

        \includegraphics[clip, trim=0 0.1in 0.3in 0.2in,
        width=0.72\textwidth]{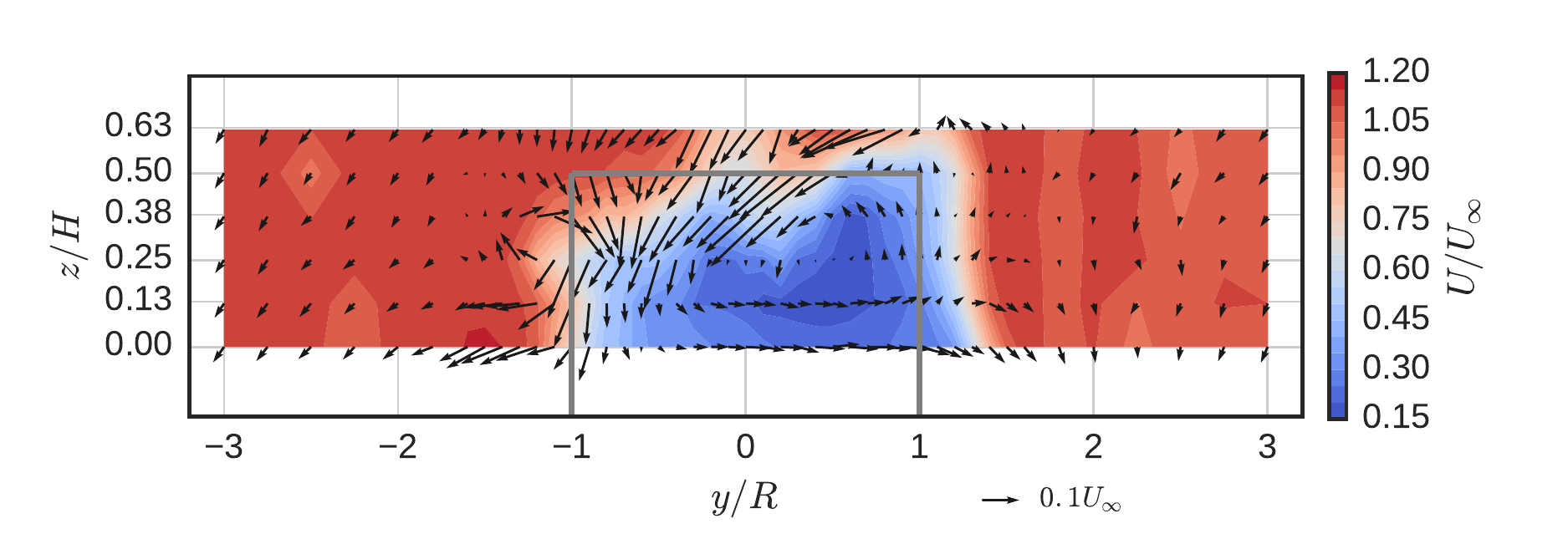}

        \caption{Mean velocity field at $x/D=1$ from experiments
            \cite{Bachant2016-RVAT-Re-dep}.}

        \label{fig:br-cfd-meancontquiv-exp}
    \end{subfigure}

    \begin{subfigure}[b]{\textwidth}
        \centering

        \includegraphics[clip, trim=0 0.2in 0 0.15in,
        width=0.8\textwidth]{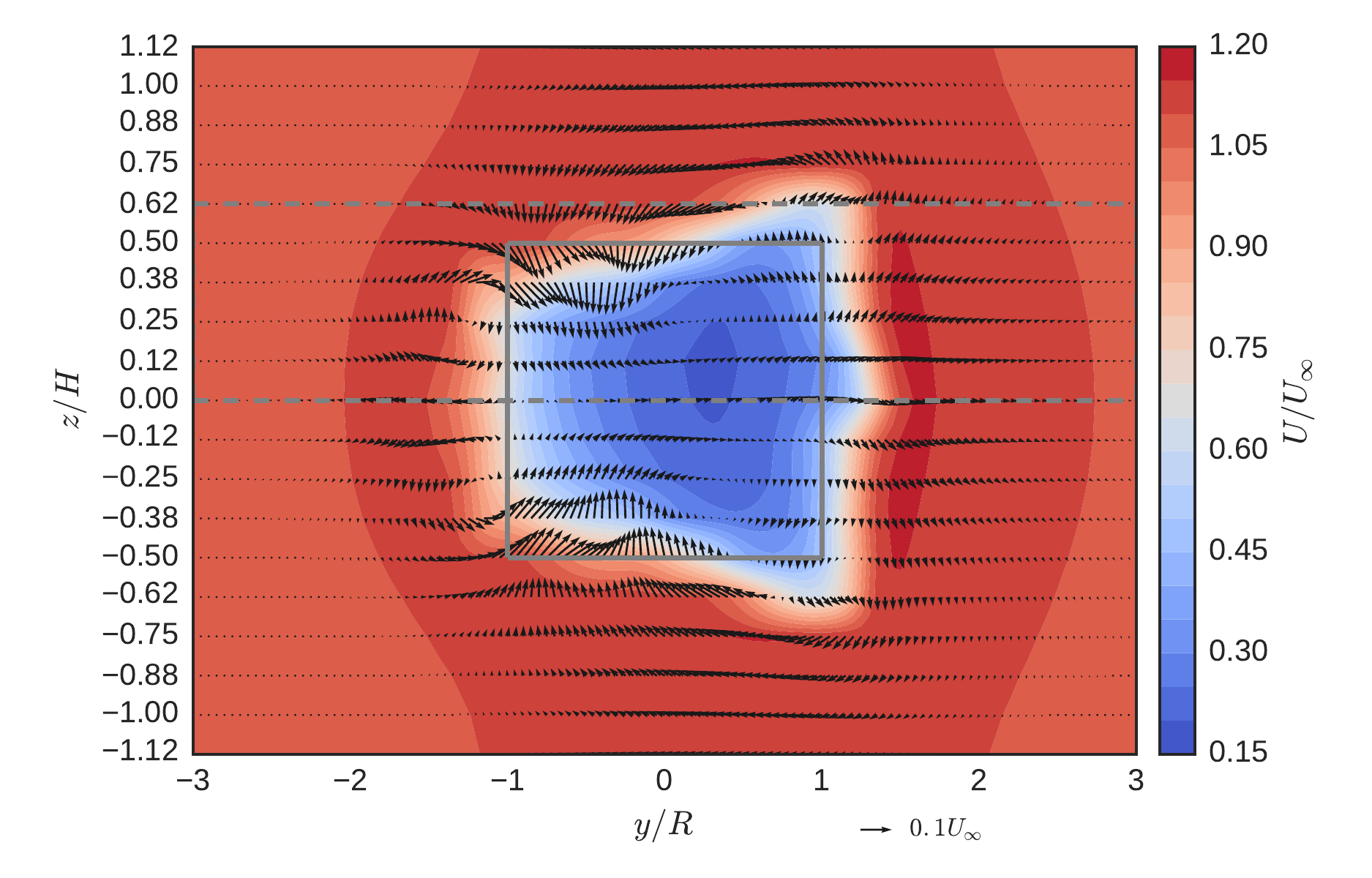}

        \caption{Mean velocity at $x/D=1$ computed by the 3-D SST model.}

        \label{fig:meancontquiv-SST}
    \end{subfigure}

    \begin{subfigure}[b]{\textwidth}
        \centering

        \includegraphics[clip, trim=0 0.2in 0 0.15in,
        width=0.8\textwidth]{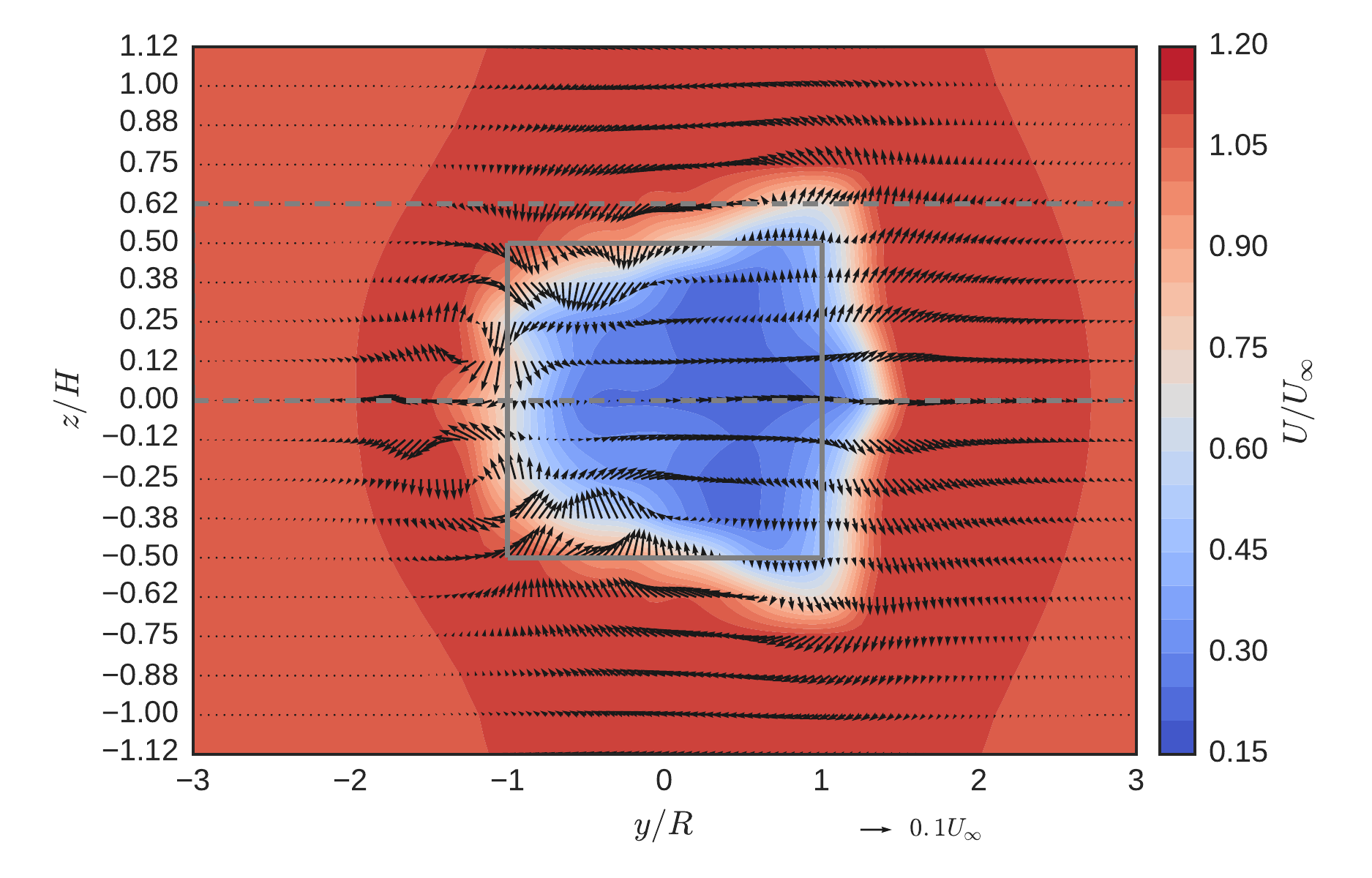}

        \caption{Mean velocity at $x/D=1$ computed by the 3-D SA model.}

        \label{fig:meancontquiv-SA}
    \end{subfigure}

    \caption{Mean velocity from experiments and 3-D CFD cases. Solid gray lines
        indicate turbine frontal area and dashed lines indicate experimental
        measurement plane.}

    \label{fig:br-cfd-mean-velocity}
\end{figure}

In order to visualize the mean velocity field, vector arrows for the mean
cross-stream and vertical components are superimposed on top of contours of the
streamwise component at $X/D=1$ in Figure~\ref{fig:br-cfd-mean-velocity}. Both
CFD models predict the general structure of the mean velocity well, though the
SA case has a slightly larger vertical mean flow component, which could be due
to stronger tip vortex generation, or lower diffusivity compared with the SST
model.

\subsubsection{Turbulence kinetic energy contours}

\begin{figure}
    \centering
    \begin{subfigure}[b]{\textwidth}
        \centering

        \includegraphics[width=0.95\textwidth]{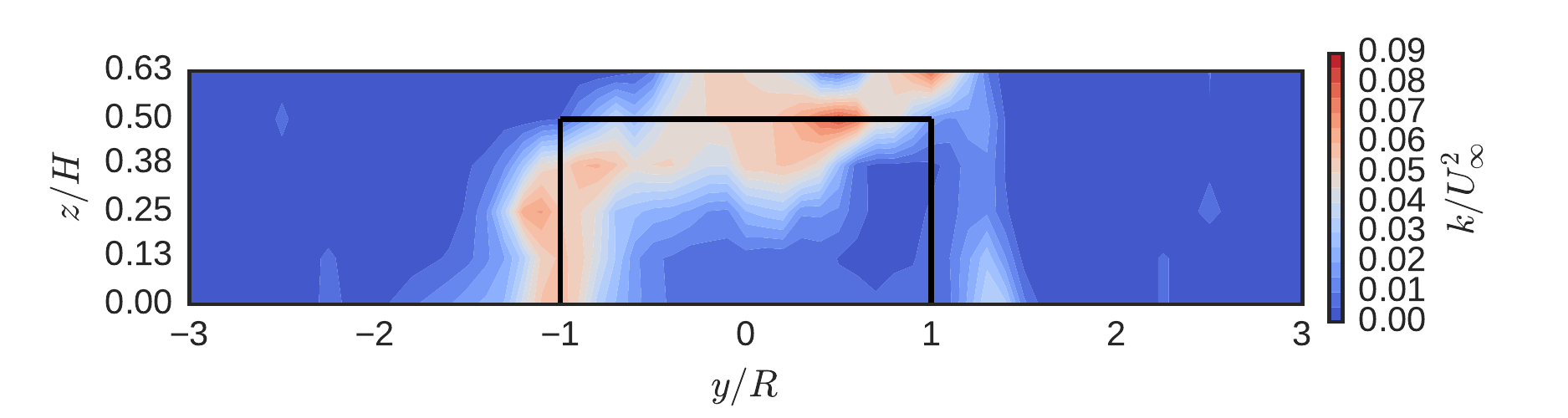}

        \caption{Turbulence kinetic energy at $x/D=1$ from experiments
            \cite{Bachant2016-RVAT-Re-dep}.}

        \label{fig:kcont-exp}
    \end{subfigure}

    \begin{subfigure}[b]{\textwidth}
        \centering

        \includegraphics[width=0.95\textwidth]{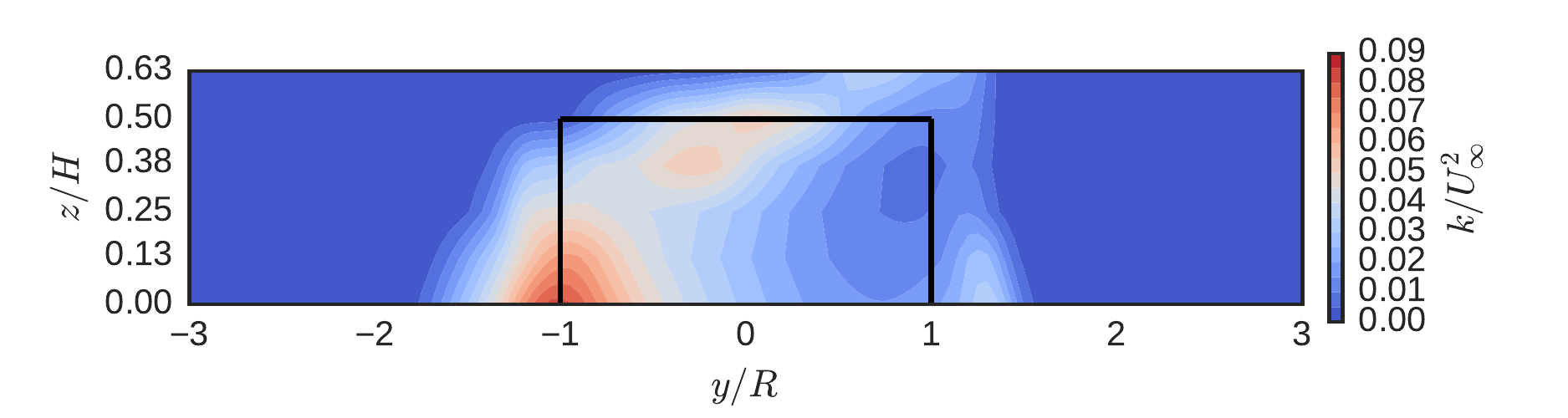}

        \caption{Turbulence kinetic energy at $x/D=1$ computed by the 3-D SST
            model.}

        \label{fig:kcont-SST}
    \end{subfigure}

    \begin{subfigure}[b]{\textwidth}
        \centering

        \includegraphics[width=0.95\textwidth]{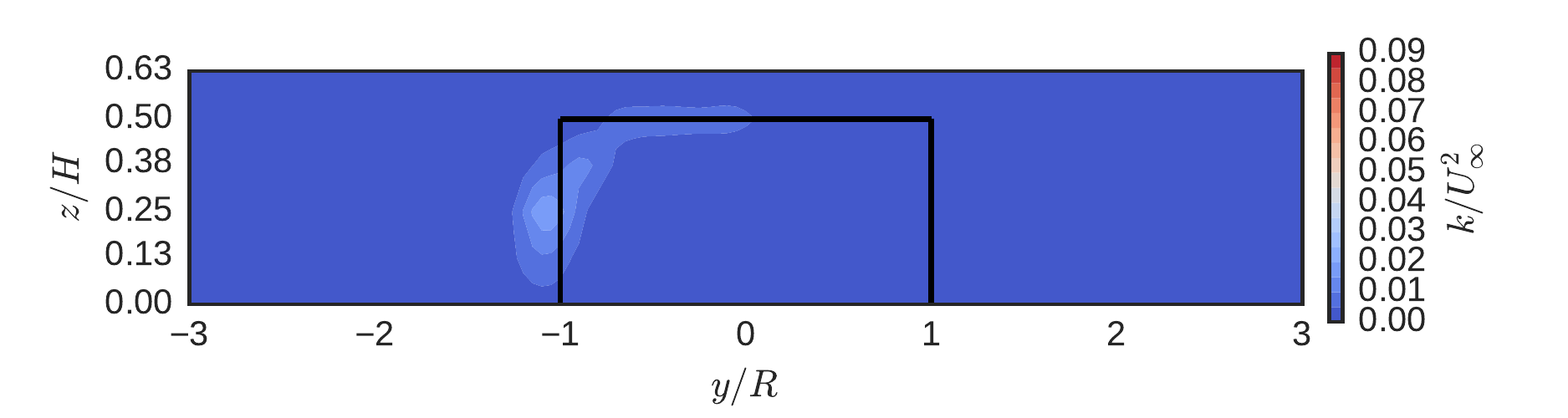}

        \caption{Turbulence kinetic energy at $x/D=1$ computed by the 3-D SA
            model.}

        \label{fig:kcont-SA}
    \end{subfigure}

    \caption{Turbulence kinetic energy from experiments and 3-D CFD cases. Solid
        black lines indicate turbine frontal area.}

    \label{fig:br-cfd-kcont}
\end{figure}

Turbulence kinetic energy contours for the experimental measurements and each
CFD case at $x/D=1$ are presented in Figure~\ref{fig:br-cfd-kcont}. As seen in
the profiles in Figure~\ref{fig:br-cfd-profiles}, the SA model was not able to
resolve the majority of the unsteadiness in the flow. In contrast, the SST model
did a good job predicting the locations and magnitudes of various peaks in $k$.
These are generated along the top of the turbine via tip vortex shedding, and
the $-y$ side of the turbine via dynamic stall. We do however see the smearing
effect from the dynamic stall vortex centered around $z/H=0$, which is likely
more of an issue with wake evolution rather than wake generation.

\subsubsection{Momentum recovery}

To get an overall idea of the wake recovery predicted by each model, we
rearrange the streamwise component of the Navier--Stokes equation to isolate
$\partial U / \partial x$---following Bachant and
Wosnik~\cite{Bachant2015-JoT}---and compute each term at $X/D = 1$ to compare
with the experimental results.

We use the RANS models' eddy viscosity to calculate the turbulent transport via
\begin{equation}
    \text{Turb. trans.} = \nu_t \nabla^2 \vec{U},
    \label{eq:turb-trans}
\end{equation}
which is a different approach from those taken on the experiments, where
Reynolds stresses were measured, but $x$-derivatives were not:
\begin{equation}
    \text{Turb. trans. (exp.)} =
    -\left(
    \frac{\partial}{\partial y} \overline{u^\prime v^\prime}
    +
    \frac{\partial}{\partial z} \overline{u^\prime w^\prime}
    \right).
\end{equation}
As such, we should not be surprised if the CFD models predict higher levels of
turbulent transport than the experiments.

\begin{figure}
    \centering

    \includegraphics[width=0.9\textwidth]{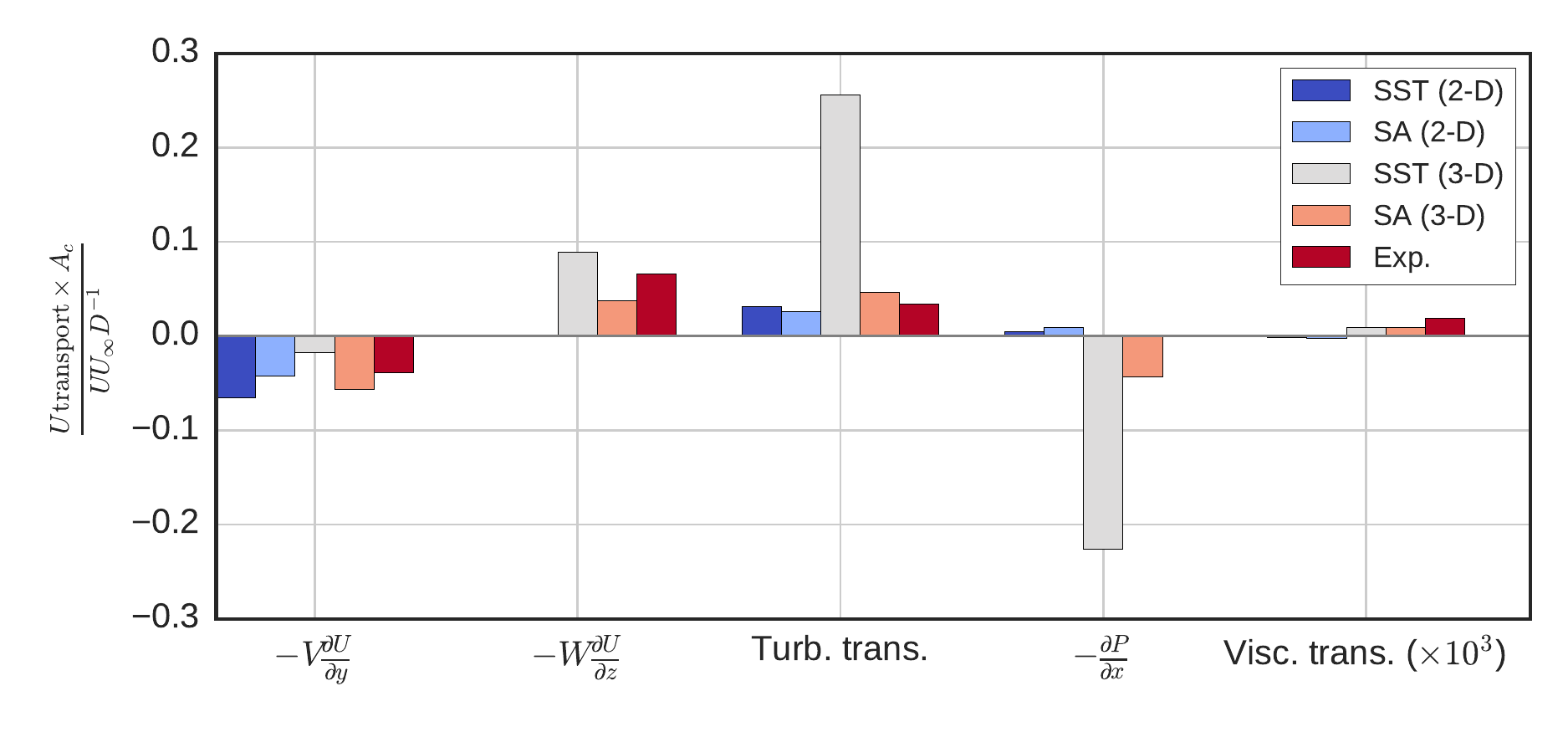}

    \caption{Weighted sum normalized momentum recovery terms for each CFD case
        and experiments\cite{Bachant2016-RVAT-Re-dep} at $x/D=1$.}

    \label{fig:br-cfd-recovery}
\end{figure}

Normalized weighted averages for each recovery term at $x/D=1$ are computed and
multiplied by the cross-sectional area of the measurement plane, or the channel
width in the 2-D cases. Results are shown in a bar chart in
Figure~\ref{fig:br-cfd-recovery}. Consistent with the relatively large Reynolds
number regime, viscous transport is essentially negligible compared with other
mechanisms. Cross-stream advection---or the tendency of streamlines to diverge
and reduce the streamwise momentum---produces a negative effect for all cases,
though the 3-D SST model predicts significantly lower values. Vertical advection
is by definition zero for the 2-D cases. The 3-D cases show varying
results---with the SST model overpredicting and SA underpredicting the vertical
velocity's effect on replenishing streamwise momentum.

Turbulent transport and streamwise pressure gradient terms show the largest
discrepancy between results. The 3-D SST case, despite doing a good job
predicting turbulence kinetic energy, significantly overpredicts the turbulent
transport term, while other CFD cases are comparable with the experiment. This
seems to be balanced by a large adverse pressure gradient, which is also present
to a smaller degree in the 3-D SA case. Interestingly, in contrast, both 2-D CFD
cases create a wake where the pressure gradient is acting to slightly accelerate
the flow at $x/D=1$. Unfortunately, pressure data were not acquired from the
experiment, though results from the 3-D delayed detached eddy simulation (DDES)
of Boudreau and Dumas \cite{Boudreau2015} concur with the adverse pressure
gradient condition.

\section{Conclusions}

A cross-flow turbine was modeled using the $k$--$\omega$ SST and
Spalart--Allmaras (SA) Reynolds-averaged Navier--Stokes turbulence models to
test their abilities to predict turbine performance and near-wake dynamics. It
was observed that when modeled in 2-D, the performance is over-predicted
compared to the data from the tow tank experiments, which was expected due to
omission of blade end effects, support strut drag, and increased blockage.
Vertical (or axial) wake dynamics were unresolved in the 2-D model, despite
being identified as a significant contributor to streamwise wake recovery, which
casts doubt on the 2-D model's applicability as a tool to study array spacing
effects.

The 3-D blade-resolved RANS simulations predicted turbine performance and
near-wake quite well, where the SA model results were closest to the
experimentally measured performance. The SA model's effectiveness gives hope for
the prospect of replacing some physical modeling with blade-resolved CFD.
However, the  overprediction of power coefficient by the SST model highlights
the level of technological and economic risk involved with CFD. That is to say,
the SST model, despite being considered relatively trustworthy in these flow
conditions, would overestimate power output by 30\%.

Both the SST and SA models did a good job predicting the mean velocity field at
one turbine diameter downstream, but the SST model was more effective at
predicting turbulence kinetic energy, since it is solved for in the turbulence
model equations. Streamwise momentum recovery terms were computed from the CFD
results over an entire cross-section of the domain at $x/D=1$. Values for
transport due to the mean pressure gradient and turbulent fluctuations varied a
lot between the two turbulence models. Both models, however, we able to at least
qualitatively resolve the vertical velocity field, which will be crucial to
predicting the performance of closely spaced CFTs.

The effectiveness at predicting the mean velocity field gives credibility to the
prospect of using the computed flow field to extrapolate the experimental
results, such that the CFD results can be used as a target for a low-order wake
generator or force parameterization. These results may also help develop new tip
loss corrections for blade element type models, which currently only exist for
axial-flow rotors, since they provide access to the pressure and shear forces
over the entire blade surface, which were not measured experimentally.
Ultimately, however, the computational cost of 3-D simulations---about 1,000 CPU
hours per simulated second (or roughly 10,000 CPU hours per operating
point)---may be too expensive to be used for CFT engineering work, especially
considering the uncertainty involved compared with physical model studies. Since
the 3-D wake results appeared to be only weakly asymmetrical in the $x$--$y$
plane about $z/H=0$, it may be possible to reduce computing power by only
modeling the top half of the rotor and using a symmetry boundary condition at
the mid-plane. However, 3-D blade-resolved RANS will likely not be practical for
turbine array simulations for quite some time, and until they are, low-order
models will need to be developed and improved for this purpose.

\begin{acknowledgments}
    The authors acknowledge funding through a National Science Foundation CAREER
    award (principal investigator Martin Wosnik, NSF 1150797, Energy for
    Sustainability, program manager Gregory L. Rorrer). The authors also thank
    Dr. Vincent S. Neary and the Sandia National Laboratories Water Power
    Program for use of their Red Mesa high performance computing cluster.
\end{acknowledgments}

% Create the reference section using BibTeX:
\bibliography{library}

\end{document}